\documentclass{emulateapj}

\slugcomment{ApJ submitted}

\shorttitle{BCG mass build up}
\shortauthors{Stott et al.}

\begin{document}


\title{The {\it XMM} Cluster Survey: The build up of stellar mass in Brightest Cluster Galaxies at high redshift}


\author{J. P. Stott\altaffilmark{1}\email{jps@astro.livjm.ac.uk}, C. A. Collins\altaffilmark{1}, M. Sahl\'{e}n\altaffilmark{2}, M. Hilton\altaffilmark{3,1},  E. Lloyd-Davies\altaffilmark{4}, D. Capozzi\altaffilmark{1}\\
M. Hosmer\altaffilmark{4}, A. R. Liddle\altaffilmark{4}, N. Mehrtens\altaffilmark{4}, C. J. Miller\altaffilmark{5}, A. K. Romer\altaffilmark{4}, S. A. Stanford\altaffilmark{6,7}, P. T. P. Viana\altaffilmark{8,9}\\
M. Davidson\altaffilmark{10}, B. Hoyle\altaffilmark{11}, S. T. Kay\altaffilmark{12}, R. C. Nichol\altaffilmark{13}}


\altaffiltext{1}{Astrophysics Research Institute, Liverpool John Moores University, Twelve Quays House, Egerton Wharf, Birkenhead CH41 1LD, UK}
\altaffiltext{2}{The Oskar Klein Centre for Cosmoparticle Physics, Department of Physics, Stockholm University, AlbaNova, SE-106 91 Stockholm, Sweden}
\altaffiltext{3}{Astrophysics and Cosmology Research Unit, School of Mathematical Sciences, University of KwaZulu-Natal, Private Bag X54001, Durban 4000, S. Africa}
\altaffiltext{4}{Astronomy Centre, University of Sussex, Falmer, Brighton, BN1 9QH, UK}
\altaffiltext{5}{Cerro-Tololo Inter-Amercian Observatory, National Optical Astronomy Observatory, 950 North Cherry Avenue, Tucson, AZ 85719, USA}
\altaffiltext{6}{University of California, Davis, CA 95616, USA}
\altaffiltext{7}{Institute of Geophysics and Planetary Physics, Lawrence Livermore National Laboratory, Livermore, CA 94551, USA}
\altaffiltext{8}{Departamento de Matem\'{a}tica Aplicada da Faculdade de Ci\^{e}ncias da Universidade do Porto, Rua do Campo Alegre, 687, 4169-007 Porto, Portugal}
\altaffiltext{9}{Centro de Astrof\'{\i}sica da Universidade do Porto, Rua das Estrelas, 4150-762 Porto, Portugal}
\altaffiltext{10}{SUPA, Institute of Astronomy, University of Edinburgh, Royal Observatory, Blackford Hill, Edinburgh, EH9 3HJ, UK}
\altaffiltext{11}{Institute for Sciences of the Cosmos (ICCUB), University of Barcelona, Marti i Franques 1, Barcelona, 08024 Spain}
\altaffiltext{12}{Jodrell Bank Centre for Astrophysics, School of Physics and Astronomy, The University of Manchester, Manchester M13 9PL, UK}
\altaffiltext{13}{ICG, University of Portsmouth, Portsmouth PO1 2EG, UK}

\begin{abstract}
We present deep $J$ and $K_{s}$ band photometry of 20 high redshift galaxy clusters between $z=0.8-1.5$, 19 of which are observed with the MOIRCS instrument on the Subaru Telescope. By using near-infrared light as a proxy for stellar mass we find the surprising result that the average stellar mass of Brightest Cluster Galaxies (BCGs) has remained constant at $\sim9\times10^{11}M_{\odot}$ since $z\sim1.5$. We investigate the effect on this result of differing star formation histories generated by three well known and independent stellar population codes and find it to be robust for reasonable, physically motivated choices of age and metallicity. By performing Monte Carlo simulations we find that the result is unaffected by any correlation between BCG mass and cluster mass in either the observed or model clusters. The large stellar masses imply that the assemblage of these galaxies took place at the same time as the initial burst of star formation. This result leads us to conclude that dry merging has had little effect on the average stellar mass of BCGs over the last $9-10$ Gyr in stark contrast to the predictions of semi-analytic models, based on the hierarchical merging of dark matter haloes, which predict a more protracted mass build up over a Hubble time. We discuss however that there is potential for reconciliation between observation and theory if there is a significant growth of material in the intracluster light over the same period. 
\end{abstract}



\section{Introduction}
Brightest Cluster Galaxies (BCGs) are the most luminous objects in the Universe in terms of stellar light and appear to be a separate population from bright ellipticals \citep{sandage1972,sandage1976,bhavsar1985,oegerle1991,postman1995,bernstein2001,bernardi2007,vonderlinden2007,vl08,lin2009}. They reside in the deep potential wells of the cores of rich galaxy clusters, thought to descend from the first regions where mass began to accumulate after the Big Bang. Their luminosities and unique environments make them ideal candidates for the testing of the mass build up in galaxies across a large fraction of the Hubble time.

A number of studies have attempted to constrain the formation epoch and evolution of BCGs by comparing their $K$ band Hubble diagram to a range of stellar population models (e.g. \citealt{abk98}; ~\citealt{cm98}; ~\citealt{nel02}). BCGs have been shown to follow passive evolution out to moderate redshifts but then differing results are seen at $z>0.5$, with some groups seeing a continuation of the passive trend (e.g. \citealt{cm98}) while others claim the high redshift BCGs are fainter and therefore less massive than their local counterparts (e.g. \citealt{abk98}). This is explained by a dependence of BCG luminosity on the mass of its host cluster \citep{bcm00}, with most studies now agreeing that the evolution of BCGs in massive clusters can be described as passive to $z\sim0.8$ \citep{stott08,whey08}. However there is still some debate over the recent merging activity of BCGs particularly in lower mass clusters and Brightest Group Galaxies with a number of studies identifying major merger candidates \citep{mulchaey2006,rines2007,tran2008}.

The favoured model for galaxy formation and evolution is via the hierarchical merging of dark matter haloes (e.g. \citealt{davis85}).  In this model the mass of a galaxy gradually increases as it merges with neighbouring systems. A major development in the field has been the advent of large cosmological simulations such as the Millennium N-body Simulation \citep{springel05} which models this hierarchical mass build up of dark matter haloes in a co-moving box 500\,$h^{-1}$\,Mpc (where $h=H_0/100\,{\rm km\, s^{-1}\,Mpc^{-1}}$) on the side. 

Semi-analytic models are commonly used to describe the complex baryonic physics of galaxies in the context of the merger histories of dark matter halos within N-body simulations (e.g. \citealt{b06}; \citealt{deluc07}). These models are an efficient way to describe the competing processes affecting baryonic matter such as those that trigger or suppress star formation i.e. gas rich galaxy merging and feedback from active galactic nuclei or supernovae. The output of observables, such as galaxy magnitudes, from the semi-analytic models has proved to be valuable for astronomers as the modelled systems can be compared directly with measurable quantities. For this reason semi-analytic models can be a useful tool for making mock catalogues, predicting the behaviour of galaxies, testing cosmological theory and assessing the feasibility of telescope observations.

One important advantage of using BCGs to study galaxy evolution is that their theoretical counterparts can be easily and unambiguously identified as the central massive galaxies in mock clusters at the same redshift as the real systems.  \cite{deluc07} presented the evolution of BCGs over a 
Hubble time, in a semi-analytic model based on the Millennium Simulation. This paper predicted that the stellar population of BCGs forms early, with 50\% of the stellar mass in place by $z=5$ and 80\% by $z=3$. However this early star formation takes place in separate components that gradually assemble into the BCGs seen in the local Universe, mainly through dry mergers that do not trigger additional star formation. So for example, at $z=1.5$ the sum of the stellar mass in all sub-components is over 90\% of the mass of the fully assembled BCG at $z=0$, whereas the stellar mass in the main progenitor is on average only 20\% of the galaxy's final mass. 

From an observational standpoint, a significant number of high redshift galaxy clusters ($z\gtrsim1$) have been discovered in recent years with X-ray surveys (e.g. \citealt{rosati2004}; \citealt{mullis05}; \citealt{stan06}; \citealt{brem06}; \citealt{lamer08}). The {\it XMM} Cluster Survey (XCS, \citealt{romer01}; \citealt{sh08}) is one such project performing a serendipitous survey to discover clusters in the XMM-Newton archive. The main goals of the XCS are to constrain cosmological parameters, measure the evolution of the hot gas through analysis of the cluster scaling relations and study galaxy evolution in clusters since the high mass cluster cores are thought to be environments comparable across all epochs. Furthermore X-ray luminosity and temperature are excellent proxies of cluster mass and enable us to directly compare the properties of real cluster galaxies with mock galaxies in similar cluster halo environments. With the advent of large wide-field optical and infrared surveys it is also possible to photometrically select high redshift galaxy clusters  based on the properties of their constituent galaxies, further increasing the number of known $z>1$ clusters (e.g. \citealt{RCS04}; \citealt{swin07}; \citealt{Eisen08}). 

In this paper we present near-infrared observations of 20 high redshift galaxy clusters ($0.8<z<1.5$), derive stellar masses for their BCGs  and compare to the latest semi-analytic models based on the Millennium N-body Simulation \citep{deluc07,springel05}. We test how robust our results are to different star formation histories produced by three independent stellar population codes: \cite{bc03}; \cite{mar2005} and BaSTI \citep{piet2004,percival2009}. This work builds on the result of \cite{collins09} in which we found that the average stellar mass of the BCGs in 5 of the highest redshift galaxy clusters is not significantly different to that of BCGs in the local Universe.  

Unless otherwise stated a Lambda Cold Dark Matter ($\Lambda$CDM) cosmology ($\Omega_{M}=$0.3, $\Omega_{Vac}=$0.7, $H_{0}=$70 km s$^{-1}$ Mpc$^{-1}$) is used throughout this work.

\section{Cluster sample}
Table \ref{tab:sample} details our sample of 20 of the most distant, spectroscopically confirmed, galaxy clusters, including the highest redshift X-ray selected cluster XMMXCS J2215.9 -- 1738 \citep{stan06,hil07, hil09}. The sample consists of clusters discovered by various X-ray surveys and several selected by optical methods that show extended X-ray emission (see Table \ref{tab:sample}). The clusters all have spectroscopically confirmed redshifts in the range $0.8\lesssim z\lesssim1.5$ and have X-ray luminosities of $ 1\lesssim L_{X}\lesssim19\times10^{44}\rm erg\, s^{-1}$. 

To anchor our analysis to the local Universe we also include a low redshift comparison sample ($0 \lesssim z\lesssim 0.3$, $0.7 \lesssim L_{X}\lesssim 20.0\times10^{44}\rm erg\, s^{-1}$) published in \cite{stott08}. This sample is a good low redshift comparison as the clusters cover a similar range in mass (average mass at $z<0.1$ is $6.8 (\pm1.5) \times 10^{14}$M$_\odot$) to the low redshift haloes of the Millennium Simulation (average mass at $z=0$ is $7.5 (\pm3.5) \times 10^{14}$M$_\odot$).

\begin{table*}
\begin{center}
\caption[]{The cluster sample}
\label{tab:sample}
\tiny\begin{tabular}{lllllllll}
\tableline\tableline
Cluster & R.A.\ & Dec.\ & $z$ & $ T_X$ & Cluster & BCG $m_{K_{s}}$&BCG &$ T_X$ reference\\
&\multicolumn{2}{c}{(J2000)}&& &Mass&&Stel. Mass\tablenotemark{a}&\\
&&&&keV&10$^{14}$M$_\odot$&&10$^{12}$M$_\odot$&\\
\tableline
\noalign{\medskip}
1. CL J0152.7-1357				&01h52m41s&-13d57m45s	&0.83	&$5.4^{+0.9}_{-0.9}$						&$4.5^{+2.7}_{-2.2}$		&16.96$\pm$0.08&0.58$\pm$0.11&\tiny{\cite{vik09}}\\
2. XLSS J022303.0 -- 043622\tablenotemark{b}		&02h23m53.9s&-04d36m22s	&1.22	&$3.5^{+0.4}_{-0.4}$		&$1.8^{+0.9}_{-0.7}$		&17.72$\pm$0.01&0.61$\pm$0.08&\tiny{\cite{brem06}}\\
3. XLSS J022400.5 -- 032526\tablenotemark{b}		&02h24m00s&-03d25m34s	&0.81	&$3.6^{+0.4}_{-0.4}$		&$2.3^{+1.4}_{-0.8}$		&16.49$\pm$0.10&0.85$\pm$0.18&\tiny{\cite{andreon2005}}\\
4. RCS J0439 -- 2904				&04h39m38s&-29d04m55s	&0.95	&$1.5^{+0.3}_{-0.2}$					&$0.5^{+0.4}_{-0.2}$		&17.70$\pm$0.08&0.40$\pm$0.07&\tiny{\cite{hicks08}}\\
5. 2XMM J083026 + 524133		&08h30m25.9s&52d41m33s	&0.99	&$8.2^{+0.9}_{-0.9}$						&$8.5^{+4.1}_{-3.4}$ 		&16.58$\pm$0.05&1.24$\pm$0.22&\tiny{\cite{lamer08}}\\
6. RX J0848.9 + 4452\tablenotemark{c}				&08h48m56.3s&44d52m16s	&1.26	&$6.2^{+1.0}_{-0.9}$		&$4.7^{+2.8}_{-2.0}$		&17.00$\pm$0.02&1.30$\pm$0.15&\tiny{\cite{bal07}}\\
7. RDCS J0910 + 5422			&09h10m44.9s&54d22m09s		&1.11		&$6.4^{+1.5}_{-1.2}$					&$5.3^{+4.1}_{-2.5}$		&17.88$\pm$0.05&0.48$\pm$0.08&\tiny{\cite{bal07}}\\
8. CL J1008.7 +5342				&10h08m42s&53d42m00s		&0.87	&$3.6^{+0.8}_{-0.6}$					&$2.2^{+1.6}_{-1.0}$		&16.42$\pm$0.08&1.06$\pm$0.21&\tiny{\cite{maughan2006}}\\
9. RX J1053.7 + 5735 (West)		&10h53m39.8s&57d35m18s	&1.14	&$4.4^{+0.3}_{-0.3}$						&$2.7^{+1.4}_{-1.0}$		&17.21$\pm$0.06&1.03$\pm$0.19&\tiny{\cite{hashimoto2004}}\\
10. MS1054.4 -- 0321				&10h57m00.2s&-03d37m27s	&0.82	&$7.8^{+1.0}_{-0.9}$					&$8.5^{+4.9}_{-3.2}$		&16.04$\pm$0.10&1.35$\pm$0.29&\tiny{\cite{branchesi2007}}\\
11. CL J1226 + 3332				&12h26m58s&33d32m54s	&0.89	&$10.6^{+1.1}_{-1.1}$					&$13.9^{+6.6}_{-5.4}$		&16.00$\pm$0.06&1.66$\pm$0.31&\tiny{\cite{maughan2004}}\\	
12. RDCS J1252.9 -- 2927			&12h52m54.4s&-29d27m17s	& 1.24	&$7.2^{+0.4}_{-0.6}$					&$6.1^{+2.3}_{-2.4}$		&17.36$\pm$0.03&0.89$\pm$0.11&\tiny{\cite{bal07}}\\
13. RDCS J1317 + 2911			&13h17m21.7s&29d11m18s	&0.81	&$4.0^{+1.3}_{-0.8}$						&$2.7^{+2.9}_{-1.3}$		&17.27$\pm$0.15&0.41$\pm$0.10&\tiny{\cite{branchesi2007}}\\
14. WARPS J1415.1 + 3612			&14h15m11.1s&36d12m03s	&1.03	&$6.2^{+0.8}_{-0.7}$					&$5.2^{+2.9}_{-1.9}$  		&16.76$\pm$0.04&1.15$\pm$0.19&\tiny{\cite{branchesi2007}}\\
15. CL J1429.0 + 4241				&14h29m06.4s&42d41m10s	&0.92	&$6.2^{+1.5}_{-1.0}$  					&$5.5^{+5.3}_{-2.0}$		&17.43$\pm0.20$&0.47$\pm$0.13&\tiny{\cite{maughan2006}}\\
16. CL J1559.1 + 6353				&15h59m06s&63d52m60s	&0.85	&$4.1^{+1.4}_{-1.0}$ 					&$2.8^{+3.2}_{-1.5}$ 		&17.21$\pm0.09$& 0.49$\pm$0.10&\tiny{\cite{maughan2006}}\\
17. CL 1604+4304					&16h04m25.2s&43d04m53s	&0.90	&$2.5^{+1.1}_{-0.7}$					&$1.2^{+1.6}_{-0.6}$		&17.61$\pm$0.09&0.38$\pm$0.07&\tiny{\cite{lubin2004}}\\
18. RCS J162009 + 2929.4			&16h20m09.4s&29d29m26s	&0.87	&$4.6^{+2.1}_{-1.1}$					&$3.4^{+5.5}_{-1.4}$		&17.63$\pm$0.12&0.35$\pm$0.07&\tiny{\cite{bignamini2008}}\\
19. XMMXCS J2215.9 -- 1738\tablenotemark{d}		&22h15m58.5s&-17d38m03s	&1.46	&$4.1^{+0.6}_{-0.9}$		&$2.1^{+1.9}_{-0.8}$		&18.72$\pm$0.01&0.39$\pm$0.05&\tiny{\cite{stan06}}\\
20. XMMU J2235.3 -- 2557			&22h35m20.6s &-25d57m42s	&1.39	&$8.6^{+1.3}_{-1.2}$					&$7.7^{+4.4}_{-3.1}$		&17.34$\pm$0.01&1.26$\pm$0.14&\tiny{\cite{ros09}}\\
\tableline
\end{tabular}
\tablenotetext{a}{The stellar mass errors quoted include the photometric error but not the uncertainty in the stellar population (see \S\ref{sec:ssp} for a full treatment).}
\tablenotetext{b}{Based on XMM archival data analysed by XCS \citep{sh08}} 
\tablenotetext{c}{Archival photometry \citep{yamada2002}.}
\tablenotetext{d}{Based on X-ray analysis to appear in Hilton et al. in prep}
\end{center}
\end{table*}

\begin{figure*}

\includegraphics[]{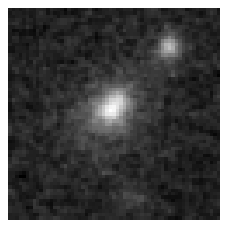}
\includegraphics[]{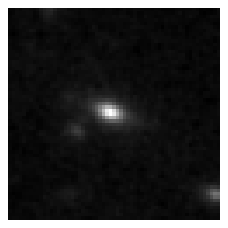}
\includegraphics[]{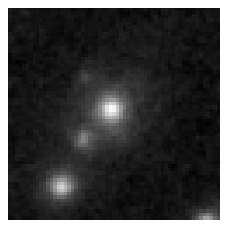}
\includegraphics[]{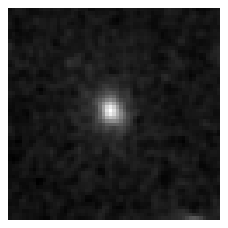}
\includegraphics[]{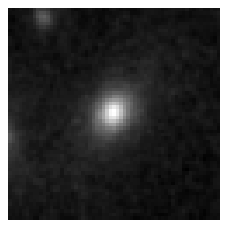}
\includegraphics[]{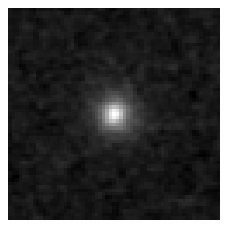}
\includegraphics[]{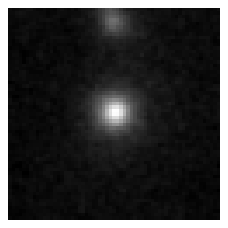}

\tiny\begin{tabular}{p{19.2mm}p{19.2mm}p{19.2mm}p{19.2mm}p{19.2mm}p{19.2mm}p{19.2mm}}
1&2&3&4&5&7&8
\end{tabular}

\includegraphics[]{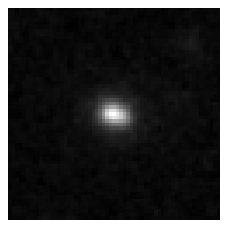}
\includegraphics[]{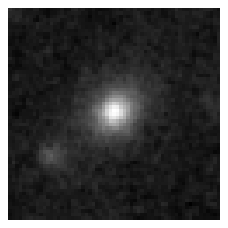}
\includegraphics[]{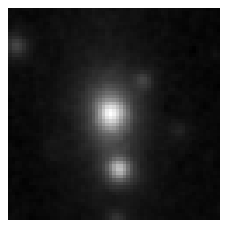}
\includegraphics[]{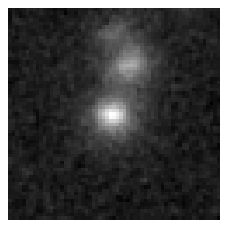}
\includegraphics[]{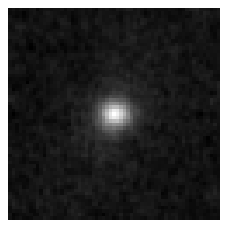}
\includegraphics[]{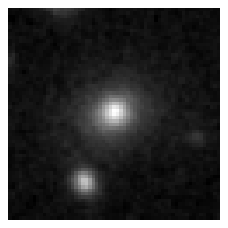}
\includegraphics[]{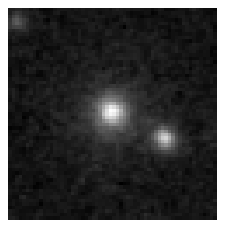}

\tiny\begin{tabular}{p{19.2mm}p{19.2mm}p{19.2mm}p{19.2mm}p{19.2mm}p{19.2mm}p{19.2mm}}
9&10&11&12&13&14&15
\end{tabular}

\includegraphics[]{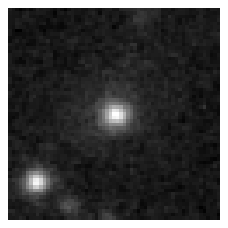}
\includegraphics[]{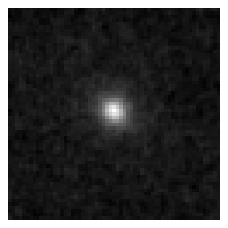}
\includegraphics[]{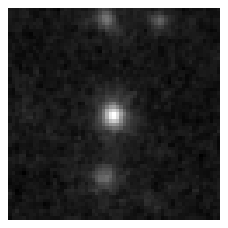}
\includegraphics[]{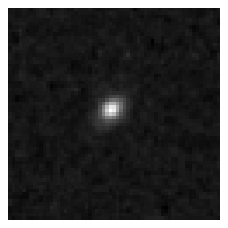}
\includegraphics[]{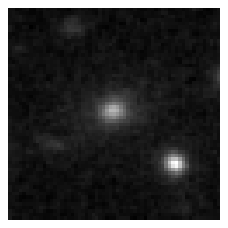}

\tiny\begin{tabular}{p{19.2mm}p{19.2mm}p{19.2mm}p{19.2mm}p{19.2mm}}
16&17&18&19&20
\end{tabular}

\caption[]{$7\times7$arcsecond postage stamp $K_s$ band images of our 19 Subaru MOIRCS observed BCGs  numbered as in Table 1. RX J0848.9 + 4452 is absent as this is an archival BCG \citep{yamada2002} for which we do not possess Subaru MOIRCS imaging.}
\label{fig:post}
\end{figure*}

\subsection{Cluster mass}
\label{sec:cmass}
A number of authors have identified  a weak correlation between BCG mass and their host cluster mass \citep{edge91,cm98,  bcm00, br08, stott08, whey08} which does not change significantly with redshift out to $z\simeq0.8$. 
Therefore in order to compare measured and  predicted BCG masses in a meaningful way it is necessary that our cluster sample is well matched to the masses of simulated clusters in the Millennium Simulation with which we are comparing. The clusters of interest from the simulation are the 125 most massive systems in the redshift snapshots $z=0.76$, $z=1.08$ and $z=1.5$,  selected for comparison with observations \citep{deluc07}. Halo masses $M_{200}$ are measured at a radius ($R_{200}$) inside which the average mass density is 200 times the critical density of the Universe. 
 
We calculate $M_{200}$ and the associated uncertainties of our sample based on the observational results for the mass--temperature ($M$--$T_X$) relation, which is preferable to the $L_X$--based determinations used in \cite{collins09} due to the putative presence of cluster cooling cores. The cluster X-ray temperatures used are listed in Table  \ref{tab:sample}. We parameterize the $M$--$T_X$ relation as
\begin{equation}
\left( \frac{M_{500}}{10^{14}   M_{\sun}} \right) = M_* \left(\frac{T_X}{1\,{\rm keV}}\right)^{\alpha} E^{\beta}(z) \,,
\end{equation}
with a log-normal scatter $N(0,\sigma_{\log T})$. Here, $E(z)$ is the standard Hubble parameter at redshift $z$. 
The $M_{500}$ masses are converted to $M_{200}$ using the standard NFW-profile prescription by \cite{HK03} with a halo concentration parameter 
$c=5$. We include a $\sigma = 10\%$ uncertainty on the normalization $M_*$ (similar to typical expected uncertainties in mass estimation, e.g. \citealt{NVK07}) along with the estimated measurement uncertainties on the temperatures, and from these derive the uncertainties on the $M_{200}$ values by a simple Monte Carlo simulation.

We use the parameter values based on the \cite{Maughan07} derived $M$--$T_X$ relation, using the ``centre excluded'' estimates in their Table 1:
$$
\log_{10}(M_*) = -0.57 \,,\\ 
\alpha = 1.72 \,,\\
\beta = -0.82 \,,\\
\sigma_{\log T} = 0.06\,.
$$

This normalization agrees with the relatively well-established value $M_{500}(z=0.05,T_{X}=5\, {\rm keV}) \approx 3 \times 10^{14} h^{-1} M_{\sun}$, e.g. \cite{PSW01,RB02,Viana03,Vik06}. The scatter is the same as in \cite{Mantz09}, which is consistent with \cite{Maughan07} although the latter does not derive the $M$--$T_X$ scatter explicitly. 

Finally we note that cluster mass estimates based on the somewhat steeper $M$--$T_X$ relation by \cite{Mantz09} give similar results.

Crucially these cluster masses cover the range of massive haloes seen in the equivalent redshift snapshots of the Millennium Simulation. The \cite{deluc07} simulated cluster samples at $z=0.76$, $z=1.08$ and $z=1.5$ have mass ranges at these redshifts of $2.4-13.6\times10^{14}\,M_\odot$, $1.5-9.8\times10^{14}\,M_\odot$ and $1.0-7.5\times10^{14}\,M_\odot$ respectively. The average mass of the combined high redshift simulated haloes is $2.6\,(\pm0.1) \times 10^{14}\,M_{\odot}$, compared to the average mass for our sample of $4.5\,(\pm0.7) \times 10^{14}\,M_{\odot}$. However, based on the known trend of BCG mass with $M_{200}$ (see \S\ref{sec:masscal}) this $60\%$ difference in average cluster mass equates at most to a shift in the average BCG mass of around $10\%$, which is within the measurement uncertainties. The consequences of this mismatch in cluster mass between the high redshift observations and the simulation is discussed in \S\ref{sec:sim}.

\section{Observations and data reduction}
The observations were taken with the MOIRCS camera \citep{ich06} on the 8.2m Subaru telescope which provides imaging and low-resolution spectroscopy over a total field-of-view of $4'\times7'$ with a pixel scale of $0.117''$ per pixel.  
This is achieved by dividing the Cassegrain focal plane and then re-focussing the light through identical optics onto two HAWAII-2 $2048 \times 2048$ CCDs, each covering $4'\times3.5'$. Observations were taken in $0.5''$ seeing on the nights of August 8th and 9th 2007 and in $0.3''$ -- $ 0.6''$ seeing on the nights of  December 16th 2008 and April 18th 2009 with the clusters centered on Detector 2. A circular 11-point dither pattern of radius $25''$ was used for both bands to ensure good sky subtraction. The modal integration times were 25 mins at $J$ and 21 mins at $K_{s}$ although we observed some of the higher redshift clusters for 50\% longer when scheduling allowed. These exposures reach a $5 \sigma$  limiting magnitude of at least $J \simeq 22.5$ and $K_{ s} \simeq 21.5$ (Vega). 

The data were reduced using the external IRAF package MCSRED. The data were flat fielded, sky subtracted, corrected for distortion caused by the camera optical design, and registered to a common pixel coordinate system. The final reduced images on which we performed the photometry were made by taking the 3$\sigma$ (s.d.) clipped mean of the  dither frames. The BCG photometry was extracted in an identical manner to the low redshift comparison sample from \cite{stott08} using SExtractor (version 2.5)  MAG\_AUTO magnitude, which is found to be within $\sim$0.1 magnitude of the total for extended objects \citep{martini2001}. As a test of this method we also performed large aperture (50 kpc) photometry on the BCGs finding the values to be consistent with those of MAG\_AUTO to within 0.05 mag. This ability to exclude light from close neighbors by using MAG\_AUTO ensures that we are not including additional flux that would bias our stellar masses to higher values. We choose a global photometric background over a local one to control for any bias introduced by a low surface brightness halo or intracluster light in the vicinity of the BCG.  
To calculate the colors of the BCGs the images are first matched for seeing using the IRAF task PSFMATCH and then we run SExtractor in dual image mode so that the $K_{s}$ band detections extract the $J$ band catalogue with identical positions and apertures to ensure accurate color determination. This photometry is performed within fixed 8 kpc circular apertures at the cluster redshift.

The photometry was calibrated to the Vega system using a combination of standard star observations and the 2MASS and UKIDSS catalogues. The typical photometric errors are 0.01 and 0.08 for the standard star and survey calibrated data respectively.

Additional archival BCG photometry is included for the cluster RX J0848.9 + 4452 \citep{yamada2002} which is a total $K_s$ magnitude measured with a large aperture and is thus comparable to our own photometric analysis.

\subsection{BCG selection}
The BCG selection for a cluster is usually obvious from visual inspection of the images as for such rich clusters they are the prominent galaxy closest to the X-ray centroid often with a cD-like profile, however we chose to formalize this by studying the tip of the red sequence in the color magnitude relation. For each cluster we identified the red sequence with $J-K_{s}$ color and selected the brightest galaxy from the $K_{s}$-band magnitudes of all the red sequence galaxies within a projected distance of 500~kpc from the cluster X-ray centroid as for approximately $95\%$ of clusters the BCG lies within this radius \citep{lm04}. The only non-obvious case is J2215.9-1738 where the object identified as the BCG is a spectroscopically confirmed member lying 300 kpc from the X-ray centroid which is only marginally brighter than several others in the cluster. A full discussion of this identification and the properties of other candidates is presented in \cite{hil09}. $K_s$ band images of our high redshift BCGs are shown in  Fig. \ref{fig:post}.

\section{Stellar mass}
\subsection{Initial stellar mass calibration}
\label{sec:masscal}
\begin{figure}
\includegraphics[scale=0.5]{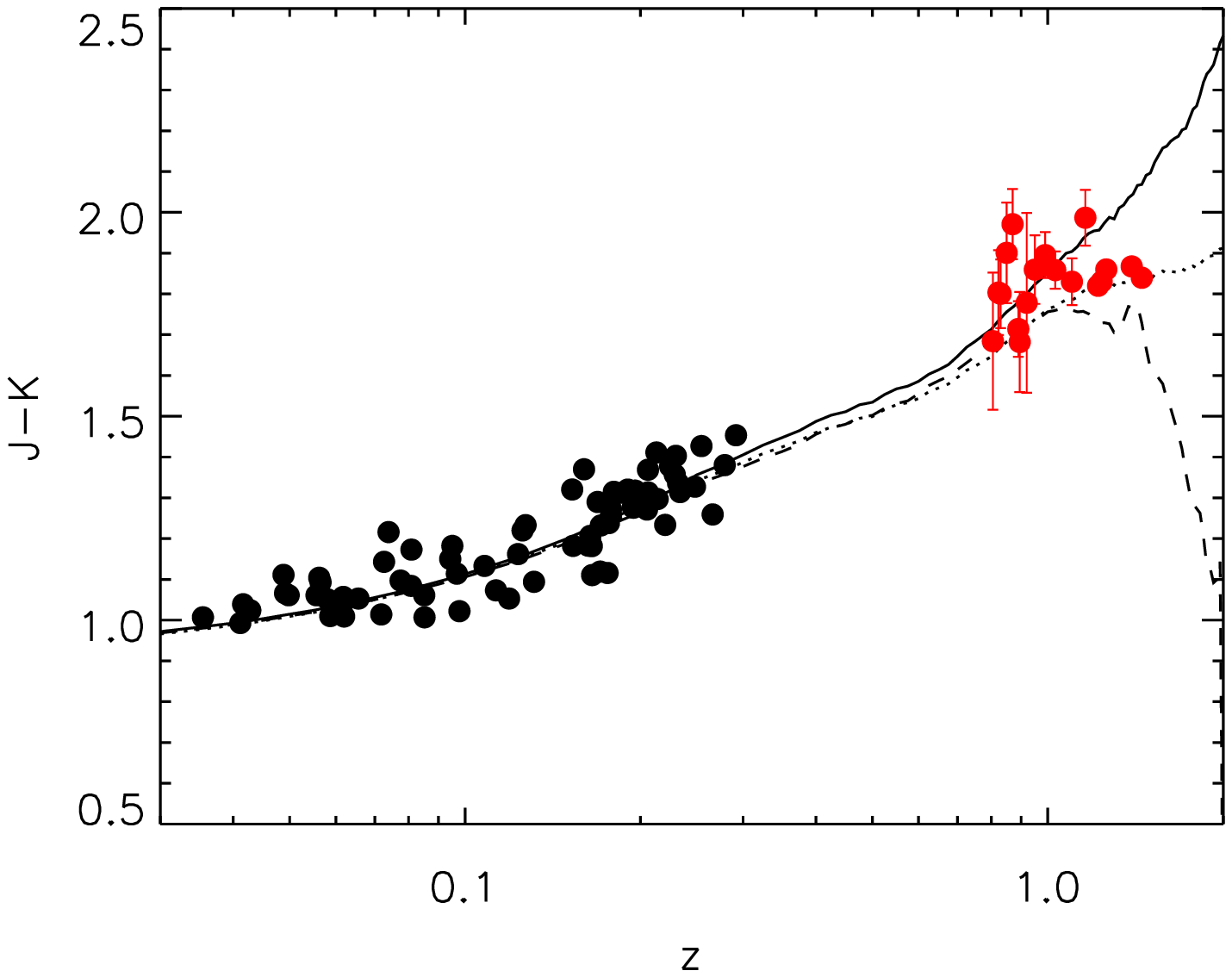}
\caption[]{$J-K_{s}$ color vs redshift for our high redshift sample (red) with 1$\sigma$ error bars and the low redshift analogue sample of \cite{stott08} (black). Two \cite{bc03} solar metallicity, simple stellar population models are included with formation redshifts $z_{f}=2$ (dashed) and $z_{f}=5$ (solid) as well as the model chosen to mimic the star formation histories of the \cite{deluc07} semi-analytic model which forms 50\% of its stars by $z=5$ and 80\% by $z=3$ (dotted).}
\label{fig:jkz}
\end{figure}

To compare the stellar mass build up in our observed BCGs with those of the semi-analytic model we need to derive stellar mass from the $K_{s}$ band luminosity of the galaxies. We first do this by calculating absolute $K_{s}$ band magnitudes for our galaxies by choosing a model appropriate to the stellar population of the modelled BCGs. The \cite{deluc07} semi-analytic model predicts a star formation history for the BCGs in which 50\% of the stars have formed by $z=5$ and 80\% by $z=3$, albeit in separate components that have not yet coalesced to form the final mass of the galaxy. To model this evolution a stellar population from \cite{bc03} with a \cite{ch03} initial mass function (IMF) is implicit in the semi-analytic model. In Fig. \ref{fig:jkz} we investigate the stellar population of our observed BCG sample by plotting the $J-K_{s}$ color against redshift and comparing to two \cite{bc03} simple stellar population (SSP) models with a Chabrier IMF and Solar metallicity. 
A composite stellar population (CSP) with an exponentially decaying star formation rate with $\tau$=0.9\,Gyr which mimics the average stellar population present in the simulation (i.e. 50\% of the stars have formed by $z=5$ and 80\% by $z=3$) is also plotted. From this plot we can see that although there is some scatter in the BCG near-infrared colors at $z\sim$1 they are consistent with the CSP model. It should be stressed that this color evolution only gives information about the stellar population (e.g. age and metallicity) and not the mass of the system. 

With the initial assumption that our observed BCGs have similar stellar populations to those of \cite{deluc07} we calculate their stellar masses using the mass-to-light ratios from the CSP and normalise these masses and the model BCG masses from \cite{deluc07} at $z=0$. We discuss the consequences of this assumption in \S\ref{sec:ssp} and compare a larger set of stellar population models appropriate to our BCGs.

\begin{figure*}
\includegraphics[scale=1]{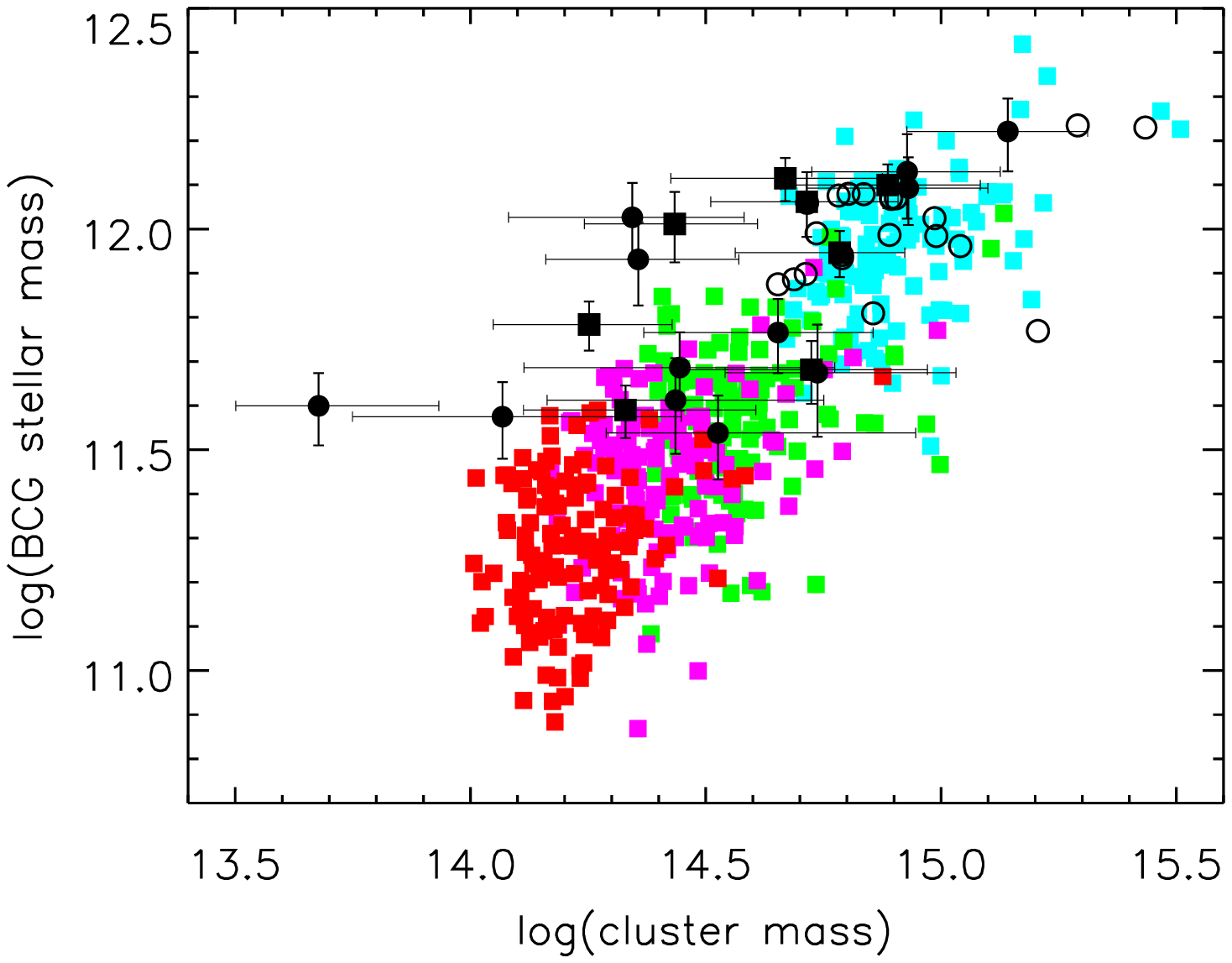}
\caption[]{BCG stellar mass ($M_{\odot}$) vs cluster mass ($M_{\odot}$) for our sample (filled black circles $z<1$ and filled black squares $z>1$ to demonstrate that there is no redshift dependence). The error bars in the cluster mass include both the X-ray observation and full mass--temperature scaling errors, whereas the stellar mass errors include the photometric errors not the uncertainty in the stellar population see \S\ref{sec:ssp} and Fig. \ref{fig:massevo}. The colored squares represent the BCGs from the model of \cite{deluc07} at the redshift snapshots $z=0$, 0.76, 1.0 and 1.5 (cyan, green, pink, red). The open circles show the $z<0.1$ BCGs from the low $z$ comparison sample of \cite{stott08}}
\label{fig:massmass}
\end{figure*}

By using this technique we find stellar masses for our BCGs in the range $\rm 3.45$ -- $\rm 16.63\times10^{11}M_{\odot}$ with a biweight scale value of $\rm 8.52(\pm1.00)\times10^{11}M_{\odot}$ which is consistent with the local BCG mass of $8.99(\pm0.82)\times10^{11}M_{\odot}$. So when considering our entire sample with stellar masses derived directly from the CSP model we find that on average there has been no significant change in the stellar mass of BCGs out to at least $z=1.5$. The corresponding stellar masses from the simulation at $z=0.76, 1.08, 1.5$ are respectively $3.84\pm0.14, 2.91\pm 0.10, 1.92\pm0.07$ in units $10^{11}M_{\odot}$. The results of this analysis can be seen in Fig. \ref{fig:massmass} in which we plot stellar mass vs cluster mass for our BCG sample (black filled) and those of  \cite{deluc07} at four different redshift snapshots (z=0, 0.76, 1.0 and 1.5 corresponding to cyan, green, pink and red squares respectively). From this plot it is clear that the average mass of the observed BCGs is significantly higher than that of the model BCGs for clusters of similar mass and redshift.

We note here that the choice of semi-analytic model does not affect our results as the `Durham models' (i.e. \citealt{b06}) also give a near identical discrepancy to that seen here \citep{collins09}.

From Fig. \ref{fig:massmass} we see that the highest mass BCG is found in the highest mass cluster which may suggest a link between stellar mass and environment as seen by \cite{edge91,cm98, br08, stott08, whey08}. However for our sample of 20 the correlation between BCG mass ($M_{\rm BCG}$) and $M_{200}$ is weak with a power law exponent of $0.42\pm0.12$ and a Spearman's rank analysis indicating this correlation is significant to the 99\% level. Determination of this correlation in the literature from larger samples at lower redshift typically show a small dependency; for example, \cite{whey08} find a power-law exponent of $0.12\pm0.03$, with a Spearman's rank correlation significant to greater than the 99.9\% level. However, as mentioned in \S\ref{sec:cmass} it is clear from Fig. \ref{fig:massmass} that there is some mismatch in cluster mass between the high redshift clusters and the corresponding simulated halos; we discuss the effect of this is \S\ref{sec:sim}.

\subsection{Dependence on stellar population models}
\label{sec:ssp}
Rather than rely on the potentially naive stellar mass calibration used in \S\ref{sec:masscal} we look now at a range of stellar population models and codes with physically motivated input parameters to study their effects on the stellar mass evolution result. For completeness we utilize three leading independent stellar population codes in this analysis namely: \cite{bc03}; \cite{mar2005} and BaSTI \citep{piet2004,percival2009}, all of which are now widely used for extragalactic astronomy. In \cite{collins09} we investigated mass-to-light ratios extensively at $z=1.3$ and found that our result held for the majority of combinations of stellar population synthesis code, age and metallicity. The notable exception to this was for young and sub-solar metallicity SSPs generated by the code of \cite{mar2005} which, because of a strong emphasis on the Asymptotic Giant Branch phase of stellar evolution, gives young stellar populations ($\sim1\,\rm Gyr$) red colors degenerate with old age and high metallicity models. 

\begin{figure}
\includegraphics[scale=0.5]{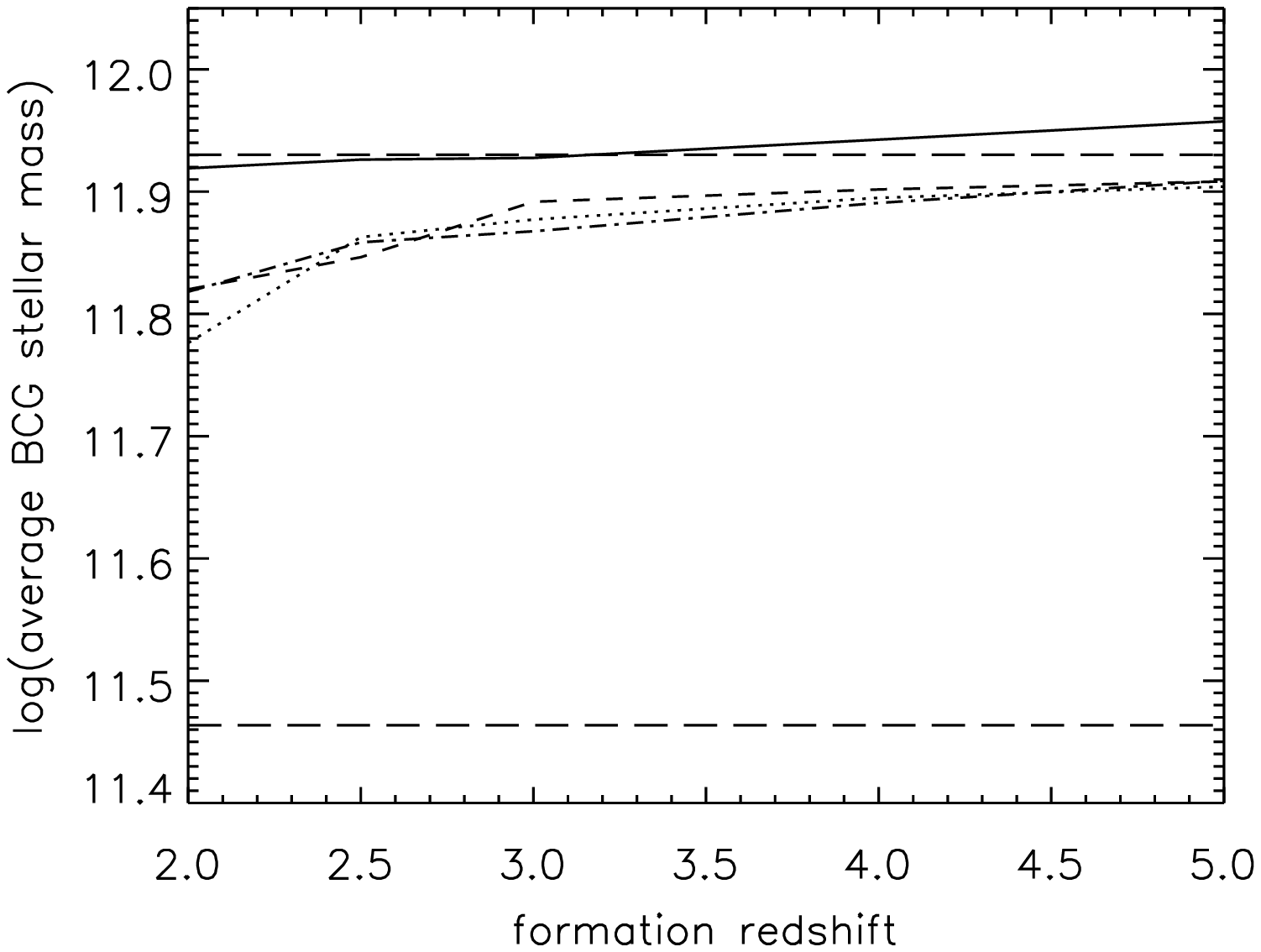}
\caption[]{Average stellar mass ($M_{\odot}$) vs formation redshift for our high redshift BCG sample when using mass-to-light ratios derived from three stellar population codes \cite{bc03}, BaSTI \citep{piet2004,percival2009} scaled solar, BaSTI $\alpha$ enhanced and \cite{mar2005} (solid, short dashed, dot dashed and dotted lines respectively) with fixed twice-solar metallicity, evaluated at the average redshift of our sample ($z=1.0$). The upper long dashed line represents the average local BCG stellar mass  $8.99\pm0.82\times10^{11}M_{\odot}$ while the lower long dashed line represents the average stellar mass of the \cite{deluc07} semi-analytic model at $z=1$ ($2.91 \pm0.10 \times10^{11}M_{\odot}$).}
\label{fig:massevo}
\end{figure}

Until recently reliable metallicity determinations were available for only a few local BCGs. However  \cite{loubser2009} examined the stellar populations of 49 BCGs in the local Universe with high signal to noise spectra, concluding that on average they have at least twice-solar metallicity ($[{\rm Z/H}]=0.31\pm0.17$) and enhanced $\alpha$ elements ($[{\rm E/Fe}]=0.41\pm0.09$), suggesting an intense burst of star formation and subsequent quiescence. 
We use this information to rule out the low metallicity models and repeat the mass-to-light ratio analysis of \cite{collins09}, concentrating on the variation in the age of the stellar component at twice-solar metallicity and including $\alpha$ enhancement (available only for BaSTI). Due to the differing metallicity sampling and IMFs available for the three stellar population codes, we use: \cite{bc03} with $\rm 2.5\,Z_{\odot}$ and Chabrier IMF \citep{ch03}; \cite{mar2005} with $\rm 2.2\,Z_{\odot}$ and Kroupa IMF \citep{kroup2001}; and BaSTI with $\rm 2.0\,Z_{\odot}$ and Kroupa IMF. We calculate the mass-to-light ratios derived from these models for galaxies at the average redshift of our sample $z=1.0$ and ages corresponding to formation redshifts $2<z_{f}<5$ \,(2.6--4.7 Gyr at $z=1$). The results of this analysis can be seen in Fig. \ref{fig:massevo}, with the \cite{bc03} models giving the highest stellar masses followed by BaSTI (scaled solar then $\alpha$ enhanced) and the \cite{mar2005} models giving the lowest. The average stellar mass predicted by the semi-analytic model at $z=1.0$ is $2.91(\pm0.10) \times10^{11}M_{\odot}$. Therefore results of the mass-to-light ratio analysis have 3$\sigma$ significance or greater for all stellar population models considered here. This confirms that our result from \S\ref{sec:masscal} is robust to the influence of reasonable, physically motivated, star formation histories generated by independent synthesis codes.

\subsection{Dependence on cluster mass}
\label{sec:sim}

Due to the nature of detecting high redshift clusters in flux limited X-ray surveys, the 
clusters in our sample are relatively high mass systems. Because of this there is some mismatch 
between the average cluster mass of our sample and the average halo masses of the 
simulation (see \S\ref{sec:cmass}). Given the relationship between cluster mass and 
BCG mass, previously discussed, this may lead to an unfair comparison. To account for 
this we perform a bootstrap simulation using the observations and the Millennium 
Simulation haloes. The details of this simulation are as follows: we select a cluster at random from our observed sample, we then pick a model BCG 
from a halo of similar mass to the observation using a random normal selection with a sigma 
equivalent to the error on the X-ray inferred cluster mass as listed in Table \ref{tab:sample}. To  account 
for the discrete nature 
of the simulated redshifts we interpolate the model BCG stellar mass to the redshift of 
the observed cluster. This is procedure is repeated 10,000 times allowing for replacement. The resulting 
distribution has an average stellar mass of $3.9\pm0.2\times10^{11}M_{\odot}$ which is $\sim5 \sigma$ away from the average mass of our sample, $\rm 8.52(\pm1.00)\times10^{11}M_{\odot}$. Thus the significance of our result does not decrease when cluster selection effects are accounted for.

\section{Summary and discussion}
We have demonstrated that the average stellar mass of BCGs in the highest redshift X-ray clusters is discrepant with those from similar mass dark matter halos in semi-analytic models based on the Millennium Simulation. Instead of the gradual build up of mass through dry merging predicted by \cite{deluc07}, our observations suggest that the stellar mass in these objects has remained unchanged over the last 9 - 10 Gyr requiring a more rapid build up of stellar mass before $z=1.5$ some $4-5$ Gyrs after the big bang.

The timescale for the mass assemblage is similar to the age of the component stars ($2-3$~Gyrs), a situation that appears to resemble classical monolithic collapse \citep{ebs62, lar74} rather than hierarchical formation. To form a galaxy of stellar mass $10^{12}\,$ M$_{\odot}$ over 4 Gyrs requires a mass deposition rate of about 250 \,M$_{\odot}$ yr$^{-1}$ and an efficient mechanism to feed the gas into the inner regions of the halo where it can form stars. Unfortunately the merging process becomes inefficient for massive galaxies because merger induced shocks lead to heating as opposed to radiative cooling of the gas \citep{binney04}. One suggestion is that the early assembly of massive galaxies at $z\geq 2$ is driven by narrow streams of dense cold gas which penetrate the shock-heated region greatly increasing the efficiency of the gas deposition and associated star formation \citep{birnboim2003, keres2005, dekel09}. 

Alternatively, a deficiency may lie in the semi-analytic treatment of the physical processes in the densest environments during early hierarchical assembly; this contention is supported by the fact that current predictions are moderately consistent with observations of the evolution of luminous red galaxies \citep{wake06, alm08}, whereas our results, which focus on the most massive subset of this population, the BCGs, differ much more from the model predictions.  

From a theoretical point of view this area of research is constantly evolving due to the challenges posed by observation and the increased availability of powerful computers. A recent high resolution simulation of a single cluster \citep{rands09} predicts that 50\% of the BCG final mass in a massive (10$^{15}$M$_{\odot}$) cluster has built up by hierarchical merging in the last 9 billion years; decreasing the discrepancy between our findings and theory. As this is for one cluster and not a full cosmological simulation on the scale of the Millennium Simulation it is difficult to know whether this improvement is due to low number statistics or a better approximation of N-body and stellar-dynamical effects.

One consequence of our result is that if merging is important at all since $z\simeq1$, the evolution of BCGs must be dominated by minor rather than major mergers, since the stellar mass appears unchanged since this time. Our observations are broadly consistent with the relatively low number of dry major mergers found at low redshift \citep{liu09} and the model predictions of \cite{khoch09} which show that only $10-20\%$ of galaxies more massive than $6.3\times10^{10}\, M_{\odot}$ have experienced dry major mergers within their last Gyr at any given redshift $z\leq1$. Numerical simulations find that the scale sizes of early galaxies can grow from dry minor merging by a factor $2-3$ since $z=1$, e.g. \cite{naab09}  and suggest that the cD--like haloes of BCGs are formed late, resulting in the relatively recent departure of BCGs from the Kormendy relation for ordinary elliptical galaxies \citep{rands09}. 

One possibility which might help reconcile observations and theory is the growth of stellar material constituting the diffuse intracluster light within the cluster cores. From dissipationless simulations of dark matter haloes, \cite{conroy2007} find that while BCGs do not appear to evolve strongly at $z<1$, the intracluster light surrounding such galaxies is growing substantially, with up to $\sim 85\%$ of the stars in the intracluster medium of present day clusters deposited at $z<1$ (see also \cite{willman2004}; \cite{murante2007}; \cite{Conroy2007b}; \cite{White2007}; \cite{Henriques2010}). This inside-out growth is broadly consistent with the dry minor merging scenario for the local elliptical galaxy population (\citealt{bezanson2009}; \citealt{hopkins2010}) required to explain their rapid size increase since $z\sim2$ \citep{vandokkum2009,vandokkum2010}.

Observationally, recent results have confirmed the overall importance of  intracluster light: \cite{rudick2006} find that the intracluster light constitutes $\sim 10\%$ of the entire  cluster stellar light; a result that appears to hold even for non-cD clusters \citep{feldmeier2004}. From surface photometry out to 300 kpc of  24 clusters at $z<0.1$, \cite{gonzalez2004} demonstrate that the outer cD component of BCGs traces the cluster potential and has $\sim 10$ times the total luminosity of  the inner elliptical profile. These results suggest that further work on the growth of stellar light in the outer parts of BCGs is required to provide a consensus.

\section{Acknowledgements}
Firstly, we would like to thank the anonymous referee for their useful comments which have improved the clarity of this paper. 

We acknowledge financial support from Liverpool John Moores University and the STFC. M.H. acknowledges support from the South African National Research Foundation. M.S. acknowledges financial support from the Swedish Research Council (VR) through the Oskar Klein Centre for Cosmoparticle Physics. P.T.P.V acknowledges financial support from FCT project PTDC/CTE-AST/64711/2006.

We thank  Gabriella De Lucia for making simulation results available to us in tabular form, Ichi Tanaka for developing the MCSRED package used to reduce the MOIRCS data, Maurizio Salaris for discussions on stellar population synthesis models and Ben Maughan for discussions on cluster masses.

This work is based in part on data collected at the Subaru Telescope, which is operated by the National Astronomical Observatory of Japan and the XMM-Newton, an ESA science mission funded by contributions from ESA member states and from NASA. 

This publication makes use of data products from the Two Micron All Sky Survey, which is a joint project of the University of Massachusetts and the Infrared Processing and Analysis Center/California Institute of Technology, funded by the National Aeronautics and Space Administration and the National Science Foundation.

This work is based in part on data obtained as part of the UKIRT Infrared Deep Sky Survey.

The Millennium Simulation databases used in this paper and the web application providing online access to them were constructed as part of the activities of the German Astrophysical Virtual Observatory.

IRAF is distributed by the National Optical Astronomy Observatories, which are operated by the Association of Universities for Research in Astronomy, Inc., under cooperative agreement with the National Science Foundation. 

\bibliographystyle{apj}
\bibliography{bcgmass_apj}




\end{document}